\input phyzzx
 \hsize=15.8cm
\vsize=23cm
\voffset=0pt

\newwrite\ffile\global\newcount\figno \global\figno=1
\def\fig{fig.~\the\figno\nfig}
\def\nfig#1{\xdef#1{fig.~\the\figno}%
\writedef{#1\leftbracket fig.\noexpand~\the\figno}%
\ifnum\figno=1\immediate\openout\ffile=figs.tmp\fi\chardef\wfile=
\ffile%
\immediate\write\ffile{\noexpand\medskip\noexpand\item{Fig.\
\the\figno. }
\REFlabeL{#1\hskip.55in}\pctsign}\global\advance\figno by1\findarg}

\parindent 25pt
\overfullrule=0pt
\tolerance=10000

\def\tr{{\rm tr}}

\def\s7{ Spin(7) }
\def\g2{ G_2 }
\def\cy4{ CY^4 }
\def\RR{${\rm R}\otimes{\rm R}~$}
\def\NSNS{${\rm NS}\otimes{\rm NS}~$}
\def\fhat{{\hat F}}

\REF\daia{J.  Dai, R.G.  Leigh and J.  Polchinski,  {\it New
connections between string theories},  Mod.  Phys.
Lett. {\bf A21} (1987) 2073.}
\REF\leigha{R.G.  Leigh, {\it Dirac--Born--Infeld action from
Dirichlet sigma model},  Mod.  Phys.  Lett. {\bf A28} (1989) 2767.}
\REF\wittena{E.  Witten, {\it Bound states of D-strings and p-branes},
hep-th/9510135, Nucl.  Phys.  {\bf B460} (1996)  335.}
\REF\townsenda{P.K.  Townsend, {\it D-branes from M-branes},
hep-th/9512062, {\it Phys.  Lett.} {\bf B373} (1996) 68.}
\REF\douglasa{M.  Douglas, {\it Branes within branes},
hep-th/9512077.}
\REF\schmidhubera{C.  Schmidhuber, {\it  D-brane actions},
hep-th/9601003.}
\REF\alwisa{  S.P.  de Alwis and K. Sato,  {\it $D$-strings and
$F$-strings from string loops}, COLO-HEP-368, hep-th/9601167.}
\REF\tseytlina{A.  Tseytlin, {\it Self-duality of Born--Infeld action
and Dirichlet 3-brane of type IIB superstring theory},
hep-th/9602064.}
\REF\greena{M.B.  Green and M. Gutperle, {\it Comments on
three-branes}, hep-th/9602077.}
\REF\polchinb{J.  Polchinski and E.  Witten, {\it Evidence for
heterotic - type I string duality}, Nucl.  Phys. {\bf B460}
(1996) 525.}
\REF\polchina{J.  Polchinski and A.  Strominger, {\it New vacua for
type IIB string theory}, hep-th/9510227.}
\REF\bergshoeffa{E.  Bergshoeff, M. de Roo, M.B. Green, G.
Papadopoulos and P.K.  Townsend, {\it Duality of type II 7-branes and
8-branes}, hep-th/9601150.}
\REF\romansa{L.  Romans, {\it Massive $N=2A$ supergravity in
 ten-dimensions},  Phys.  Lett. {\bf 169B} (1986) 374.}
\REF\bergshoeffb{E.  Bergshoeff and M.  de Roo, {\it D-branes and
T-duality}, hep-th/9603123.}
\REF\bergshoeffc{E.  Bergshoeff, C.M.  Hull and T.  Ort\'{\i}n, {\it
Duality in type II superstring effective action},  Nucl. Phys.
{\bf B451} (1995) 547.}
\REF\bachasa{C.  Bachas, {\it D-brane dynamics}, hep-th/9511043, Phys.  Lett {\bf  374B} (1996) 37.}
\REF\polchinc{J.  Polchinski, S.  Chaudhuri and C.V.  Johnson, {\it
Notes on D-branes}, hep-th/9602052.}
\REF\alvareza{E.  Alvarez, J.L.F.  Barb\'on and T.  Borlaf, {\it
T-duality for open strings}, hep-th/9603089.}
\REF\jackiwa{S.  Deser, R.  Jackiw and S.  Templeton, {\it
Three-dimensional massive gauge theories},  Phys.  Rev.  Lett. {\bf
48} (1982) 975.}
\REF\gunaydina{M. G\"unaydin, G.  Sierra and P.K.
Townsend, {\it
Quantization of the gauge coupling constant in a five-dimensional
Yang--Mills/Einstein supergravity theory}, Phys.  Rev.  Lett. {\bf 53}
(1984) 322.}
\REF\lougha{M. O'Loughlin, {\it Chern--Simons from Dirichlet 2-brane
instantons},
EFI-96-02, hep-th/9601179.}

\line{\hfill RU-96-20}
\line{\hfill DAMTP/96-41 }
\line{\hfill QMW-PH-96-6}
\line{\hfill hep-th/9604119}
\vskip 1cm

\centerline{D-BRANE WESS-ZUMINO ACTIONS, T-DUALITY}
\centerline{ AND THE  COSMOLOGICAL CONSTANT}

\vskip 1cm

 \centerline{ Michael B.  Green,\foot{ DAMTP,
Silver Street,
Cambridge CB3 9EW, UK\ \ \
M.B.Green@damtp.cam.ac.uk}  Christopher M.  Hull\foot{Physics  
Department, QMW,
Mile End Road, London E1 4NS, UK \ \ \
C.M.Hull@qmw.ac.uk} and Paul K.  Townsend\foot{ DAMTP,
Silver Street,
Cambridge CB3 9EW, UK \ \ \
P.K.Townsend@damtp.cam.ac.uk}}
\vskip 0.2 cm
\centerline{ Department of Physics and Astronomy,}
\centerline{ Rutgers University, Piscataway NJ
08855-0849 }

\vskip 2.0cm

\noindent
A geometrical formulation of the T-duality rules for type II
superstring Ramond--Ramond fields is presented.  This is used to  
derive the
Wess-Zumino terms in superstring D-brane
actions, including terms proportional to the mass parameter of the IIA
theory,  thereby completing partial results in the literature.
For  non-abelian world-volume gauge groups the massive type IIA
D-brane actions contain non-abelian Chern--Simons terms for the
Born--Infeld gauge potential, implying a
quantization of the IIA cosmological constant.

\vfill\eject

\chapter{Introduction}

The bosonic sector of the effective world-volume action for a type IIA
superstring Dirichlet $p$-brane in a bosonic type II supergravity
background takes the form
$$I = I_{DBI} + I_{WZ}.\eqn\worldact$$
The first term is the Dirac--Born--Infeld (DBI) action
[\daia-\greena],
$$I_{DBI} = - T_p \int\!  d^{p+1}\xi\  e^{-\phi} \sqrt {-\det
\left(g_{ij} - B_{ij}
+{\alpha' \over 2\pi} F_{ij}\right)}\ ,\eqn\dbin$$
where $\phi$, $g_{ij}$ and $B_{ij}$ are the pullbacks to the
world-volume of the Neveu--Schwarz-Neveu--Schwarz (\NSNS) supergravity
fields and $F=dV$, where $V$ is the Born--Infeld $1$-form $U(1)$ gauge
field.  The constant $T_p$ is the $p$-volume tension with mass dimension
$p+1$.  The second term in \worldact\ is the \lq Wess-Zumino' (WZ)  term
describing the coupling of the $D$-brane to the background
Ramond--Ramond (\RR) fields.  We can assemble these fields into the
complex of  differential forms,
$$C = \sum_{r=0}^9 C^{(r)}, \eqn\forms$$
where $C^{(r)}$ is a differential form of degree $r$. The fields
$C^{(r)}$ are the \RR gauge potentials of either IIA ($r$ odd) or IIB
($r$ even) supergravity.  The 9-form potential is optional because its
equations of motion force the dual of its field strength to be a
constant, $m$ [\polchinb-\bergshoeffa].  One can set $m=0$ but $m\ne
0$ is also possible, in
which case the background
fields are those of massive IIA supergravity [\romansa].

 When
$C^{(9)}=0$, i.e. $m=0$, one has [\douglasa],
$$I_{WZ}^{(p+1)} = T_p \int_{W_{p+1}}  \! C e^{({\alpha'\over 2\pi}F
-B)}, \eqn\totalact$$
where it is to be understood that one selects the $(p+1)$-form in the
expansion of the integral and that all forms in space-time  are pulled
back to the $(p+1)$-dimensional world-volume, $W_{p+1}$.  One purpose
of this paper is to generalize \totalact\ to the $m\ne 0$ case,
thereby generalizing previous partial results along these lines
[\bergshoeffb].

It is convenient to define
$$\fhat = F- {2\pi \over \alpha'} B,\eqn\fhatdef$$
where $B$ is again the pullback to the world-volume of the
corresponding space-time two-form.  In the
generalization from a $U(1)$ gauge field to  a $U(n)$ gauge field,  
which is
appropriate for $n$ coincident $D$-branes [\wittena], the DBI
world-volume  1-form
gauge potential $V$ takes values in the Lie algebra of $U(n)$ and its
Lie-algebra valued field strength $F$ is given by
$$F = dV + V^2.  \eqn\fielddef$$
In this case we can still define $\fhat$ as in \fhatdef\ if $B$ is
assumed to take values in the Lie algebra of $U(1) \subset U(n)$.

The WZ term \totalact\ has the following straightforward
generalization to the $U(n)$ case [\douglasa],
$$I_{WZ}^{(p+1)} =T_p \int_{W_{p+1}} C\  \tr
\left(e^{\alpha'\fhat/2\pi} \right),
\eqn\nonabwz$$
where the trace is taken in the $n$-dimensional representation of
$U(n)$.  The DBI term will require more extensive modification, but in
this paper we shall be concerned exclusively with the WZ term.

It should be noted that $C$ includes both the standard \RR\ potentials
of IIA and IIB supergravity in its form expansion {\it and} their
duals.  It is consistent to introduce both a potential and its dual  
if the
background fields are on shell because the Bianchi identities and the
field equations are then on the same footing.  The $C^{(9)}$ gauge  
potential is
exceptional in a number of respects.  For example, it has no dual
potential.  Its field strength does have a dual field
strength  but the field equations restrict this to be  a
constant, $m$.  This constant, which  has dimensions of mass, is
essentially the square root of the cosmological constant appearing in
the \lq massive' IIA supergravity theory [\romansa].  Non-zero
$C^{(9)}$ is therefore equivalent to non-zero $m$
[\polchinb -\bergshoeffa].  In such a background the WZ term \nonabwz\
requires $m$-dependent modifications.

These modifications have been found for $p=0, 2$ in the $U(1)$ case
[\bergshoeffb].  One result of this paper will be to extend these
results to all even $p$ and to gauge groups $U(n)$.  Specifically, we
will show that when $m\ne 0$ the integrand of \nonabwz\ is replaced
(up to a total derivative) by
$$L = C \tr\left(e^{\alpha'\fhat/2\pi} \right) + m \sum_{r=0} {1 \over
(r+1)!} \omega^{(0)}_{2r+1}, \eqn\fullwz$$
where $\omega^{(0)}_{2r+1}$ is a Chern-Simons $(2r+1)$-form with  
the property
that
$$d \omega^{(0)}_{2r+1}  = \tr\left({\alpha' F\over 2\pi}
\right)^{r+1}.\eqn\omdef$$
The form expansion of $L$ gives the WZ lagrangians used to
construct $I^{(p+1)}_{WZ}$.

Since the $m$-dependent term in \fullwz\ contributes only odd forms in
the form expansion of $L$,  only the $D$-brane actions for {\it
even} $p$ (i.e. those of the type IIA theory) acquire $m$-dependent
corrections.  This is expected because the constant $m$ can be
considered as a \RR field of the type IIA theory, equivalent to
$C^{(9)}$.  One way to see that these $m$-dependent terms must be
present in the IIA D-brane actions is by T-duality with the type IIB
D-brane actions.  This was shown for $p\le 3$ and with a $U(1)$ gauge
group in  [\bergshoeffb] using
the \lq massive' T-duality rules of [\bergshoeffa] (which complement
the \lq massless' rules given in  [\bergshoeffc]).   We will
generalize this procedure to establish that  the $m$-dependent term in
\fullwz\ is precisely that required by T-duality.  This result also
completes previous results [\bachasa-\alvareza] on T-duality of
D-brane actions for $m=0$.    However, since  the only known supersymmetric 
backgrounds
that solve the IIA field equations require the presence of an eight-brane the
D-brane actions for $m\ne 0$  should be interpreted as describing the dynamics 
of the D-brane in the presence of an eight-brane.  In such a background there 
can be additional world-volume fields arising from open strings connecting 
the D-brane to the eight-brane.  These will not be taken into account in this paper.

With a non-abelian $U(n)$ gauge group the $m$-dependent terms in
\fullwz\ are non-abelian Chern-Simons (CS) forms.  As described in  
a different
context in [\jackiwa,\gunaydina], quantum consistency in such cases  
requires a
quantization of the CS coefficient which, in this case, is the mass
parameter $m$.  Thus, an important consequence of the results of this
paper is the quantization of the IIA superstring cosmological
constant, as argued previously [\polchina,\bergshoeffa] for quite
different reasons.  This will be discussed further at the end of  
this article.

\chapter{D-brane actions in \RR backgrounds}

In order to explain the form  of the   D-brane WZ actions it will prove
very useful to first provide a uniform formulation of the \RR and \NSNS
gauge symmetries. In this and the next section we shall set
$\alpha'=1$ for
convenience.

The differential forms of even degree appearing in \forms\ are the \RR
gauge potentials of IIB supergravity while those of odd degree are the
\RR\ potentials of IIA supergravity.  These \RR\ fields are subject to
the gauge transformation,
$$\delta_\Lambda C = d \Lambda -  H \wedge \Lambda,\eqn\inva$$
where $H=dB$ is the \NSNS\ $3$-form field strength and
$$\Lambda = \sum_{r=0}^8 \Lambda^{(r)}.\eqn\lamsum$$
A further \RR\ transformation is
$$\delta_\lambda C = \lambda e^B,\eqn\conslam$$
for {\it constant} $\lambda$.  It will be convenient for our purposes
to combine the \RR\ transformations into
$$\delta_{RR} C = d \Lambda - H\wedge \Lambda + \lambda e^B.
\eqn\comgauge$$
These transformations encapsulate the local \RR\ symmetry of the  
IIA and IIB
supergravity lagrangians.  The specific choice of field variables, and
hence of the transformations, has been chosen in order to clarify the
invariances of the    D-brane WZ  actions.  Specifically,
$$\delta_{RR}  L = d \left[\Lambda \tr
\left( e^{\fhat/2\pi}
\right)  +  \lambda \sum_{r=0} {1 \over
(r+1)!} \omega^{(0)}_{2r+1}     \right], \eqn\csinv$$
which means that the WZ action \nonabwz\ changes by a surface term.
An intriguing feature of this result is that
$$\delta_{RR}L\big|_{\Lambda = C,\ \lambda=m} = dL,\eqn\cohom$$
so that the $(p+1)$-form lagrangians in the expansion of $L$ are
related by a type of cohomological descent.

We also need to consider the transformation of the background \RR
fields under the \NSNS gauge transformation $\delta_\eta B = d \eta$.
When $m=0$ the \RR potentials are invariant, $\delta_\eta C =0$, but
when $m\ne 0$ they transform non-trivially [\bergshoeffa].  In terms
of the field
definitions in this paper these transformations are,
$$\delta_\eta B = d\eta, \qquad \delta_\eta C = -me^B\eta.\eqn\etatrans$$
It is clear from the form of the WZ lagrangian when $m=0$ that it  
is invariant
under the $\eta$ gauge transformation provided that
$$\delta_\eta V = 2\pi \eta,\eqn\vtrans$$
where $\eta$ is considered to take values in the Lie algebra $U(1)
\subset U(n)$. The D-brane lagrangian for $m\ne 0$ can be found from
the requirement of $\eta$ invariance as we shall see later.
 Note that
$$[\delta_\eta,\delta_\lambda]= \delta_\Lambda |_{\Lambda = \lambda\eta
e^B},\eqn\commtrans$$
with all the other commutators vanishing, so that the combined
$\eta,\lambda,\Lambda$ transformations form a closed algebra.

The \NSNS field strength $H=dB$ is obviously invariant under all the
above transformations.  So is the field strength of the \RR\ fields,
$$R(C) = dC - H\wedge C + m e^B.\eqn\rrfield$$
The form expansion of $R(C)$ yields all the \lq modified' field
strengths of the IIA and IIB potentials {\it and} their duals.  To
relate the potentials to their duals we first define
$$R_< = \sum_{r=0}^5 R^{(r)}, \qquad R_> = \sum_{r=5}^{10} R^{(r)},
\eqn\defrg$$
and then impose the constraint,
$$R_> = * R_<,\eqn\selfd$$
where $*$ is the Hodge dual in ten dimensions.  This relates the
potentials of the background supergravity theory  to their duals and also
imposes  self-duality of the IIB five-form field strength.
Also, note that \selfd\ implies that the ten-form field strength is
such that
$$* R^{(10)} = m,\eqn\tenfield$$
which is appropriate when $m\ne 0$.\foot{The corresponding  relation in
[\bergshoeffa] looks more complicated when $B\ne 0$ but must be
equivalent after field redefinitions.  This suggests that the field
definitions used here might significantly simplify the massive IIA
supergravity.}
The Bianchi identity takes the form
$$d R(C) - H\wedge R = d\left[ R(C) e^{-B}\right] = 0.\eqn\bianchi$$

The $m=0$ WZ action \nonabwz\ can be expressed as
an integral of a covariant expression in $p+2$ dimensions on a
manifold $M_{p+2}$ with  boundary $W_{p+1}$,
$$I_{WZ}^{(p+1)} = T_p \int_{M_{p+2}}  R(C)  \tr\left(e^{\fhat/2\pi}
\right).\eqn\covacti$$
It is natural to suppose that this action remains valid when $m\ne 0$
because there is no other candidate integrand that is both gauge
invariant and reduces to the one known to be correct when $m=0$.  When
$m\ne 0$ the integrand of \covacti\ can be written as $dL$ where
$L$ is as given in \fullwz.   As  the addition of any closed
form to $L$ yields the same expression for $dL$ there is an
intrinsic ambiguity.  We will find it convenient to use the
alternative, but equivalent, lagrangian
$$L_{WZ} = \left[ n  C +  \omega R(C)\right] e^{-B},\eqn\intpart$$
where $\omega$ is defined by the requirement that
$$d \omega  = \tr\left( e^{F/2\pi} -1 \right).\eqn\omegdef$$
It is obvious from the construction that the lagrangian \intpart\
is $\eta$-invariant up  to a total derivative, and a calculation shows
 that $\delta_\eta L_{WZ} = - d\left(\eta C e^{-B}\right)$.
Also, $L_{WZ}$ varies by a total derivative under the gauge
transformation $\delta_\chi V = D\chi$.  The action is, however, not
necessarily invariant under  \lq large' gauge transformations, as will
be discussed in the last section of this paper.

We shall now show that the WZ lagrangians contained in the form expansion
of \intpart\ are related by T-duality.

\chapter{T-duality of D-brane WZ actions}

T-duality of Dirichlet $p$-brane actions amounts to the statement that
the double  dimensional reduction of the $p$-brane action yields the
\lq direct'
reduction of the $(p-1)$-brane action if the background fields of the
two actions are related by T-duality.  We shall therefore begin by
discussing T-duality for the background fields.

 T-duality presupposes the existence of a $U(1)$ symmetry of the
background.  We can choose cordinates such that the associated Killing
vector field  is $\partial/\partial y$.  In this case, the metric,  
dilaton
and all field strengths are independent of $y$.  In addition, we will
make the simplifying assumption that
$$i_yH = 0,\eqn\hassum$$
where $i_y$ ($\equiv * dy*$) indicates contraction with  $\partial
\over \partial y$.  With this assumption $H$ is invariant under
T-duality, as can be seen from the T-duality rules of [\bergshoeffc],
and this greatly simplifies the  discussion of the T-duality  
transformation
of the
WZ term.

The T-duality transformations of the \RR field strengths (implicit in
the T-duality rules of [\bergshoeffc,\bergshoeffa]) are given by
$$R(C) \to dy \wedge R(C) + i_y R(C),\eqn\tmapofr$$
where $\to$ means that the fields appearing on the left are replaced
by those on the right.  Since $(dy + i_y)^2 = 1$, T-duality is an
invertible  $Z_2$
transformation between the IIA and IIB fields.  The map \tmapofr\ of the
field strengths is induced by
$$C \to -(dy \wedge C + i_y  C), \eqn\qina$$
provided that
$${\cal L}_y C =  (m - m dy)e^B,\eqn\liedef$$
where ${\cal L}_y $ is the Lie derivative  with respect to
$\partial /\partial y$.
This implies, in particular, that
$$C^{(0)} \leftrightarrow - C^{(1)}_y,\eqn\parti$$
and \liedef\ determines $C^{(0)}$ and $C^{(1)}_y$ to be of the
form,
$$C^{(0)} = my + \bar C^{(0)}, \qquad C^{(1)}_y = - my + \bar C_y^{(1)},
\eqn\asdef$$
where $\bar C^{(0)}$ and $\bar C_y^{(1)}$ are independent of $y$, in
agreement with [\bergshoeffa].  Despite the $y$-dependence of the
potential $C$ when $m\ne 0$, the field strength $R(C)$ is
$y$-independent since
$${\cal L}_y R(C) =0.\eqn\lieonr$$

We shall now compare the double-dimensional reduction of  the
$p$-brane WZ
action with the direct reduction of the  $(p-1)$-brane WZ action.  The
double-dimensional reduction proceeds as follows.  We write
$$L_{WZ} = L^+_{WZ} + L^-_{WZ}, \eqn\csplit$$
where $L^\pm_{WZ}$ are the projections of $L_{WZ}$ defined by
$$L^+ = d\sigma\wedge i_\sigma L, \qquad L^- = i_\sigma
(d\sigma \wedge L),\eqn\pmdef$$
for arbitrary world-volume form $L$, where $\sigma$ is a particular
world-volume coordinate.  The $(p+1)$-dimensional world-volume
$W_{p+1}$ will be taken to be of the form $W_p \times S^1$, where
$\sigma$ is the $S^1$ coordinate.  If the  $p$-brane is now wrapped
around the $S^1$ factor of space-time, with coordinate $y$, then the two
$S^1$ coordinates can be identified,
$$dy = d\sigma.\eqn\identi$$
Moreover, only the $L_{WZ}^+$ projection of $L_{WZ}$ contributes to
the integral over $W_{p+1}$.  We shall now work towards a convenient
expression for $L^+_{WZ}$.

We first observe that the world-volume one-form gauge potential $V$ can
be written as
$$V = d\sigma V_\sigma + V^-,\eqn\onefor$$
and hence
$$F = - d\sigma \wedge DV_\sigma + F^-,\eqn\fdecom$$
where $DV_\sigma = d V_\sigma + [V,V_\sigma]$.  To establish the
T-duality map between the D-brane actions in the non-abelian case
($n>1$) it will be necessary  to restrict $V_\sigma$ to the Lie
algebra of the $U(1)$ subgroup of $U(n)$.  We therefore set
$$V_\sigma = -  2\pi  \Phi,\eqn\dualco$$
where $\Phi$ is a scalar world-volume field.
If we take the compactification radius to be $l$ so that $\sigma$ is
identified with $\sigma + 2\pi l$ (using $dy=d\sigma$), then a $U(1)$
transformation of $V$ with group element $e^{i\sigma/l}$ shifts $\Phi$
to $\Phi + 2\pi \tilde l$, where $l\tilde l = 4\pi^2$.
The world-volume field $\Phi$ is therefore a map
from the $S^1$ factor of space-time in the T-dual theory to a
$(p-1)$-brane world-volume.
Using \dualco\ we have  $DV_\sigma = - 2\pi d\Phi$, and hence, from
\fdecom\
$$\tr\left( e^{F/2\pi}\right) = \left[ \tr\left(e^{F/2\pi}\right)
\right]^- + d\sigma
\wedge d\Phi \left( e^{F/2\pi}\right) ,\eqn\decfr$$
from which it follows that
$$\omega^{+} = d\sigma \wedge d\Phi\wedge \omega - n d\sigma\ \Phi\
.\eqn\omdecom$$

Space-time forms can be decomposed in the same manner as \csplit, so
that,
$$R(C)  = R^+(C) + R^-(C), \eqn\splitr$$
where
$$R^+(C) = dy\wedge i_y R(C), \qquad R^- (C) = i_y(dy \wedge
R(C)).\eqn\rpmdef$$
In view of the relation $dy = d\sigma$, the $\pm$ components
of space-time forms pull back to $\pm$ world-volume forms.  Noting
also that $H = H^-$ by virtue of the assumption \hassum, we have
$$L_{WZ}^+ = C^+ n e^{-B}  + \omega R^+ e^{-B} + \omega^{+} R^-
e^{-B}.\eqn\supeq$$
Substituting  the expression \omdecom\ into the last term gives
$$ \omega^{+}R^- (C) e^{-B}  = - d\sigma\wedge \left[
\omega d \Phi R^- e^{-B} - n d\Phi C^- e^{-B}  \right]
   - n d\sigma\wedge d \left[\Phi C e^{-B} \right].\eqn\refg$$
The last term in this expression will give a surface term in the integral
over $W_p$  and can be ignored.  Thus,
$$\int_{W_{p+1}} L_{WZ} = \int_{W_{p+1}} L^+_{WZ} = \int_{W_p \times
S^1}\left[n \left(C^+ + d\sigma d\Phi
C^-\right) +
\left(\omega R^+ - d\sigma \omega d \Phi R^-
\right)\right] e^{-B}.\eqn\csag$$
We could now perform the $\sigma$
integral to get the double dimensional reduction of $L_{WZ}$,  which
would then have to be rewritten using
the T-duality rules.  It is more convenient to reverse the order of
these steps, making use of the fact that the T-duality transformations
of the components of $C$ and  $R(C)$ are,
$$\eqalign{& C^+ \to - dy \wedge C^+, \qquad C^- \to -i_y C^-, \cr
&R^+(C) \to dy \wedge R(C), \qquad R^-(C) \to i_y
R(C).\cr}\eqn\rdet$$
This yields, on setting $dy =d\sigma$,
$$\int_{W_p \times S^1} L_{WZ} \to -  \int_{S^1}d\sigma \int_{W_p}
\left[n (C^- + d\Phi i_y  C) + \omega^-
(R^- + d\Phi i_y R) \right] e^{-B}.\eqn\lcsto$$

Before attempting to integrate over $\sigma$ we should determine
whether the integrand is $\sigma$-dependent.  This possibility arises
when $m\ne 0$ because in that case $C$ is $y$-dependent.  In fact,
using \liedef,
$$C e^{-B} = \bar C e^{-B} + my - mydy,\eqn\consc$$
where $\bar C$ is $y$-independent.  Thus,
$$(C^- + d\Phi i_y C)e^{-B} = (\bar C^- + d\Phi i_y \bar C) e^{-B} +
my - my d\Phi.\eqn\rescm$$
The $my$ term contributes a constant to the zero-brane lagrangian, and
may be ignored, while the $my d\Phi$ term contributes a total
derivative (proportional to $m d\Phi$) to the one-brane lagrangian, which
may also be ignored.   Effectively, therefore, the $W_p$ integral in
\lcsto\ is $\sigma$-independent so performing the $\sigma$
integral yields
$$\int_{W_p} L_{WZ} \to - 2\pi l\int_{W_p} \left[n (C^- + d\Phi i_y C)
+ \omega^-
(R^- + d\Phi i_y R) \right] e^{-B}.\eqn\lcsnew$$
 If $\Phi$ is now interpreted as the $S^1$
space-time coordinate then a form such as $C^- + d\Phi i_y C$ is just the
decomposition into $\mp$ projections of the $D=10$ form $C$, i.e.  
it is the
\lq direct' reduction of $C$.  Furthermore, the form $\omega^-$ on
$W_{p+1}$ is just $\omega$ on $W_p$ since $W_p$ has no $\sigma$
coordinate.
We conclude that the integral on the
right-hand side of \lcsnew\ is just $L_{WZ}$,
so
$$T_p \int_{W_{p+1}} L_{WZ} \to -  2\pi lT_p \int_{W_p}  
L_{WZ}.\eqn\resag$$
This result shows that the D-brane actions are related by T-duality
provided that\foot{The sign is irrelevant since the tensions should be
identified with the absolute values  of the WZ coefficients.}
$$T_{p-1} = (2\pi l_p) T_p,\eqn\tense$$
where $l_p$ is the radius of the compact dimension in the $D=10$ theory
with the Dirichlet $p$-brane;  that is, $l_p=l_A$ for $p$ even and
$l_p=l_B$ for $p$ odd, where $l_A$ and $l_B$ are related by $l_A l_B
= 4\pi^2 $.
It should be remarked that had we used \fullwz\ instead of \intpart\
then it would have been necessary to consider boundary terms in the
action to establish T-duality.  This accounts for the ambiguity in the
approach of [\bergshoeffb] in which, in effect, boundary terms are
ignored.

  For later  considerations it will be important to
appreciate that the constants $T_p$ are not the physical tensions;
these are
$$T_p^{phys} = {T_p\over g},\eqn\physten$$
where $g$ is the string coupling constant.  Under $T$-duality $g\to
2\pi l g$, so the physical tension of the $(p-1)$-brane
found by double dimensional reduction is
$$T_{p-1}^{phys} =  T_p^{phys}.\eqn\tenph$$

\chapter{Quantization of the cosmological constant}
When $m\ne 0$ the WZ lagrangian for even $p$ contains the term
$$I_p(V) = mT_{2r}  {1\over (r+1)!} \int_{W_{2r +1}}
\omega^{(0)}_{2r+1}, \qquad (p=2r),\eqn\menters$$
which is a Chern-Simons term for the {\it world-volume} $U(n)$  
gauge field.
As is well known, under a gauge transformation $V\to g^{-1} Vg +
g^{-1} dg$ the action \menters\ changes by a term proportional to the
winding number of  $g(\xi)$.  If $W_{p+1}$ is the $(p+1)$-sphere
then the maps $g$ are classified by $\pi_{p+1}(G)$, for a gauge group
$G$.  For $G = U(n)$ one has $\pi_{2p+1} \supset Z$.  Thus, for
sufficiently large $n$
there are always large $U(n)$ gauge transformations for which the
action is not invariant.  As shown in [\jackiwa] single valuedness of
$e^{iI_p(V)}$, required for quantum consistency of the
world-volume field theory, implies a quantization of the coefficient
of  $I_p(V)$.
The resulting quantization condition in our case is
$$m T_p (\alpha')^{1+p/2} = 2\pi \nu,\qquad (p \ {\rm even}),\eqn\quan$$
for integer $\nu$.  This is actually a series of quantization
conditions, one for each even value of $p\le 8$.  The consistency of
these relations requires (for $p$ even and a given $\nu$)
$$\alpha' T_p =  T_{p-2},\eqn\tcons$$
which can be shown to be satisfied (for any $p$) by iteration of
\tense\ and use of
 the relation
$$ l_A l_B = 4\pi^2 \alpha'.\eqn\rell$$
The relations \tcons\ should not be confused with  the
well-known quantization
condition on the products $T_pT_{6-p}$.

The physical tensions $T^{phys}_p= T_p/g$ should be independent of  
the radius.
Since the coupling $g$ is not invariant under T-duality we shall set
$g=g_A$ or $g=g_B$, according to which of the two type II theories we
are considering.   Consider first the IIB theory.   The condition that
$T^{phys}_p$ is independent of
the compactification radius implies that
$$T_p =c g_B (\alpha')^{-(p+1)/2}, \qquad (p\ {\rm
odd}),\eqn\twobv$$
for some  dimensionless constant $c$ that is independent of $g_B$
(and is  determined by normalization conventions in the string theory).
 One can
now use this in  \tense\ to obtain an  expression for $T_p$ when $p$
is even.  Substitution of  this  into \quan\ gives
$$m = (c g_B l_B)^{-1} \nu.\eqn\mres$$
If $ g_B$ is set to unity the quantization condition $m \propto \nu  
/l_B$ of
[\bergshoeffa] is recovered.  The result in [\bergshoeffa] was
obtained by requiring consistency between T-duality and $SL(2;Z)$
U-duality of type IIB theory.  However, the factor of $g_B^{-1}$ is
important for the purpose of rewriting \mres\ in  IIA
terms since $g_B l_B = 2\pi \sqrt{\alpha'} g_A$.  Thus, the mass
quantization condition may be written entirely in terms of the IIA
theory as
$$m ={1\over  2\pi  c g_A {\sqrt{\alpha'} }}\nu.\eqn\endm$$
The same result follows more directly from the condition  that the  
physical
D-brane tensions
 of the IIA theory be independent of $l_A$.  Thus the mass
scale set by the quantization condition of $m$ is not a new scale in
the theory but is the same as the one set by the zero-brane mass.

We will conclude with some comments on the $p=2$ case, which is of
particular interest  since it describes a membrane
that is supposed to descend from the membrane of eleven-dimensional
M-theory [\townsenda,\schmidhubera].  Let
$$V =  V_0 + \tilde V,\eqn\vdfe$$
where $V_0$ is a $U(1)$ gauge potential and $\tilde V$ an $SU(n)$
gauge potential.  Then $I_p(V)$ can be written as\foot{A CS term for the
euclidean
$p=2$ D-brane has appeared previously in a different context [\lougha].}
$${m T_2 \over 2} \int_{W_3}  V_0 d V_0 + {m T_2 \over 2} \int_{W_3}
\tr\left(\tilde Vd\tilde V + {2\over 3}\tilde  V^3 \right).
\eqn\csfor$$
The first term is the topological mass term found in [\bergshoeffb].
As pointed out in
[\bergshoeffb] this term prevents the dualization of $V_0$ to
a world-volume scalar and hence appears to obstruct the $D=11$
interpretation of the
Dirichlet two-brane.
Indeed, the mass of the type IIA theory has no known 11-dimensional
interpretation, and this
  is an interesting challenge to
the idea that all superstring theories should be unified in $D=11$
M-theory.  The quantization condition \endm\ might help resolve  
this puzzle
since $m$ is quantized in units of $1/R_{11}$, where $R_{11}$ is the
radius of the eleventh dimension.   This unit therefore goes to zero
in the decompactification limit.

\vskip 0.3cm
{\it Acknowledgments}

We are grateful to the Rutgers Physics Department for its
hospitality.  We are also grateful to Michael Douglas,
 George Papadopoulos,  Ashoke  Sen, and  Andy
Strominger  for helpful comments.

\refout

\bye